\documentclass[runningheads]{llncs}

\usepackage[T1]{fontenc}
\usepackage{graphicx}
\usepackage{diagbox}
\usepackage{amsfonts}
\usepackage{multirow}
\usepackage{enumitem}
\usepackage{hyperref}

\usepackage{xcolor,soul}
\hypersetup{
    colorlinks,
    linkcolor={blue!50!black},
    citecolor={blue!50!black},
    urlcolor={blue!50!black}
}
\usepackage{color}

\usepackage{bbm}
\usepackage{amsmath}
\usepackage[capitalise]{cleveref}

\usepackage{float}
\usepackage{color, colortbl}
\definecolor{Gray}{gray}{0.88}
\definecolor{LightCyan}{rgb}{0.8,1,1}
\definecolor{LightGreen}{rgb}{0.0, 0.5, 0.0}
\definecolor{LightRed}{rgb}{1, 0.7, 0.7}
\definecolor{mydarkgreen}{RGB}{0,160,0}

\begin{document}
%

\title{Continual Domain Incremental Learning for Privacy-aware Digital Pathology}



%
\author{Pratibha Kumari\inst{1}\textsuperscript{$\star$}\textsuperscript{$@$}\orcidID{0000-0003-3681-3700}\and
Daniel Reisenb\"uchler\inst{1}\textsuperscript{$\star$}\orcidID{0000-0001-8820-0530
}\and 
Lucas Luttner\inst{1}\orcidID{0000-0003-1103-9077}\and 
Nadine S. Schaadt\inst{2}\orcidID{0000-0001-7685-8087} \and 
Friedrich Feuerhake\inst{2}\orcidID{0000-0002-1234-982X} \and 
Dorit Merhof\inst{1,3}\orcidID{0000-0002-1672-2185} 
}
\authorrunning{Kumari et al.}
%
\institute{Faculty of Informatics and Data Science, University of Regensburg, Regensburg, Germany 
\and
Institute of Pathology, Hannover Medical School, Hannover, Germany 
\and 
Fraunhofer Institute for Digital Medicine MEVIS, Bremen, Germany\\
\textsuperscript{$\star$}Equal contribution, 
\textsuperscript{$@$}Correspondence (Pratibha.Kumari@ur.de)
} 

\maketitle              
%
\begin{abstract}
In recent years, there has been remarkable progress in the field of digital pathology, driven by the ability to model complex tissue patterns using advanced deep-learning algorithms. However, the robustness of these models is often severely compromised in the presence of data shifts (e.g., different stains, organs, centers, etc.). Alternatively, continual learning (CL) techniques aim to reduce the forgetting of past data when learning new data with distributional shift conditions. Specifically, rehearsal-based CL techniques, which store some past data in a buffer and then replay it with new data, have proven effective in medical image analysis tasks. However, privacy concerns arise as these approaches store past data, prompting the development of our novel Generative Latent Replay-based CL (GLRCL) approach. GLRCL captures the previous distribution through Gaussian Mixture Models instead of storing past samples, which are then utilized to generate features and perform latent replay with new data. We systematically evaluate our proposed framework under different shift conditions in histopathology data, including stain and organ shift. Our approach significantly outperforms popular buffer-free CL approaches and performs similarly to rehearsal-based CL approaches that require large buffers causing serious privacy violations.

\end{abstract}
%
%

\begin{figure}[!ht]
    \centering
    \includegraphics[scale=0.4]{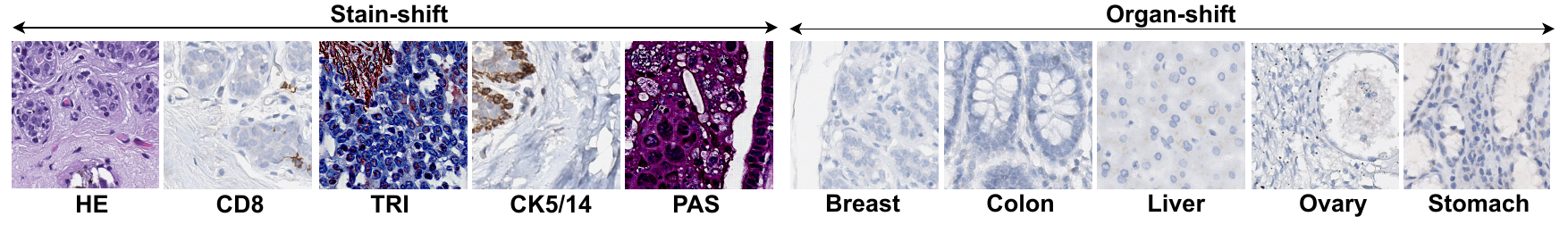}
    \caption{Example of domain-shift in digital pathology. \textit{Left}: different stainings of breast samples. \textit{Right}: Hematoxylin samples of different organs for tumor detection.}
    \label{fig:shiftExampleBoth}
\end{figure}
\section{Introduction}\label{sec:intro}
Rapid advancements in deep learning models have revolutionized digital pathology. However, such models are mostly validated in stationary environments where data is presumed to follow a static distribution, which is usually not the case in clinical settings. 
Histopathology slides originating from different organs, staining protocols, centers, etc. possess various levels of covariance shifts~\cite{guan2021domain} (see examples in Fig.~\ref{fig:shiftExampleBoth}). However, deep learning models would show a drastic performance degradation on datasets that do not follow the data distribution initially used for training the model~\cite{karani2018lifelong}. On the other hand, naively updating the model with training data of a new domain may cause disruption in the previously learned domain, also known as catastrophic forgetting~\cite{kirkpatrick2017overcoming}.
One possible solution to prevent catastrophic forgetting is to retrain the deployed model on accumulated datasets of both past and new domains. 
However, sharing past datasets is often not feasible with medical data due to privacy concerns~\cite{ravishankar2019feature}. Also, past data may no longer be accessible, which would also hamper federated learning approaches. Further, such approaches demand considerable amounts of computation time and memory to store datasets and would require full model retraining each time distribution shifts in the dataset are encountered.

Alternatively, Continual Learning (CL) has emerged as a promising incremental learning paradigm to avoid catastrophic forgetting~\cite{derakhshani2022lifelonger}. CL sequentially accumulates knowledge over a stream of datasets, frequently referred to as tasks, each with possible shifts, without truly revisiting the previous task. There are various CL strategies detailed in~\cite{kumari2023continual}, all aiming to reduce catastrophic forgetting and tackle performance drops. Medical imaging researchers have also started exploring these CL strategies for various medical image analysis applications that are prone to encounter domain shifts and novel classes, including segmentation, disease classification, drug discovery, etc.~\cite{kumari2023continual}. 
In digital pathology, Bandi et al.~\cite{bandi2023continual} provide a comparative study of existing CL strategies on cancer detection datasets, reflecting shifts among three organs (breast, colon, and head-neck). Similarly, Kaustaban et al.~\cite{kaustaban2022characterizing} simulate five domain incremental datasets by changing stain composition in a colon cancer dataset (CRC) and benchmark existing CL approaches. Some other works~\cite{derakhshani2022lifelonger,yang2023few} benchmark popular CL approaches on less complex datasets such as MedMNIST. 
Further CL research in digital pathology indicates that rehearsal-based strategies that require storing some past samples in a memory buffer tend to perform better compared to others~\cite{bandi2023continual,kaustaban2022characterizing}. However, storing past samples may cause privacy violations~\cite{holub2023privacy}, which can be a major bottleneck of such CL approaches in medical applications. A less concerning direction, storing features instead of actual images, refereed as latent-replay (e.g., chest x-ray classification~\cite{srivastava2021continual}, Ultrasound cardiac view classification~\cite{ravishankar2019feature}) is also explored.  
Besides the promising results from current research, such a strategy is compromised by complexities in sharing features across hospitals and large buffer requirements.

To address these limitations related to the sharing of medical data in histo\-pathol\-o\-gy, we propose a novel, buffer-free CL approach that takes advantage of a Gaussian Mixture Model (GMM) to encapsulate the traits of past training domains, enabling feature generation for incorporation in subsequent training sessions. We acquire real-world histological slides from different environments and curate domain-incremental scenarios to show shifts in terms of stain, organ, center, or a mix of those. An extensive evaluation is performed on three different domain shift problems in digital pathology, i.e., (1) breast tumor detection across different histological stains, (2) tumor detection across different organs, and (3) tumor detection in the presence of heterogeneous types of shifts. For comparison, we consider established buffer-free and buffer-based CL approaches.

\begin{figure}[!ht]
    \centering
    \includegraphics[scale=0.55]
    {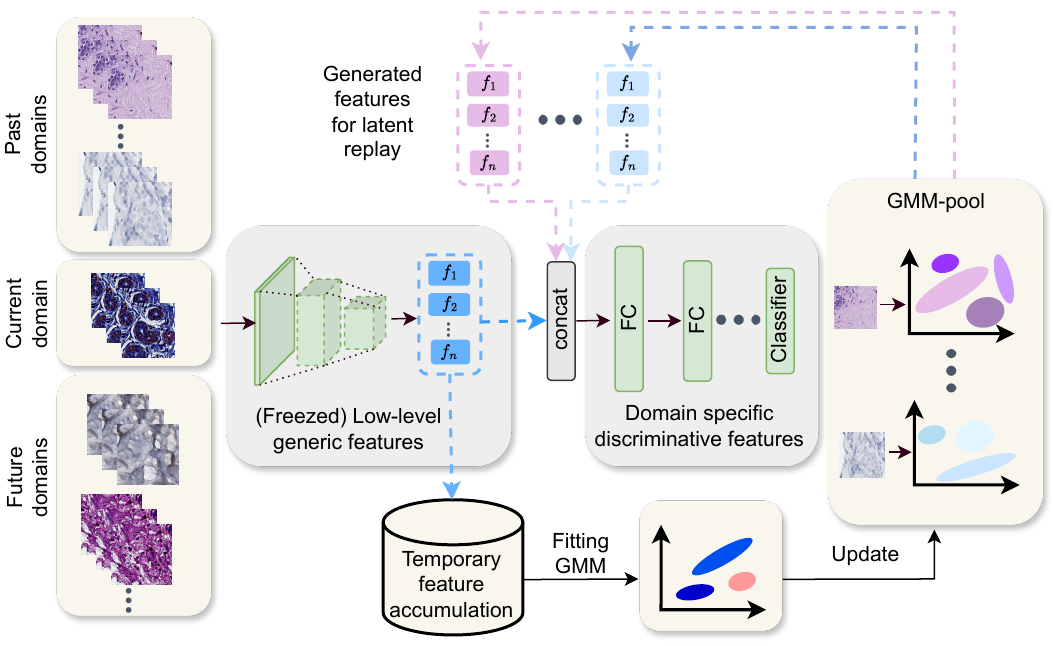}
    \caption{Overview of proposed CL framework for digital pathology}
    \label{fig:framework}
\end{figure}

\section{Method}\label{sec:proposed}
In the following, we provide the problem formulation and introduce the two key novelties of our proposed CL framework: (a) generative latent replay and (b) domain specific generators. Fig.~\ref{fig:framework} serves as overview of our proposed framework.

\textbf{Problem formulation:} The goal of GLRCL is to learn sequentially from datasets containing domain shifts and without storing any samples from past datasets while maintaining performance over previous datasets. Such an incremental learning setting where the classification task is fixed, but domain shift is observed over time, is referred to as continual learning under domain incremental scenarios. 
Each dataset in the sequence is termed as a CL \textbf{task} here. 
Formally, there is a sequence of tasks \{$T_1$, $T_2$, ....\}, where each task $T_t$ has training $D_t$ and evaluation $E_t$ set for tumor classification. Here, the training set $D_t$ = \{$x_t$, $y_t$\} contains $x_t$ patches extracted from Whole Slide Images (WSIs) and corresponding $y_t$ annotations for $\{1, .... C\}$ classes. The training data of $t^{th}$ task offers samples for learning in the current session ($t^{th}$ domain) only and is not accessible in past or future training sessions. Meanwhile, the testing data from each task is available throughout the sessions. The CL classification model $\mathbb{M}_{t}(\cdot)$ is initialized with model parameters $\theta_{t-1}$ learned during the immediate past session. $\mathbb{M}_{t}(\cdot)$ then aims to mitigate catastrophic forgetting, by learning the model parameters $\theta_{t}$ that minimize the loss ($\textit{\textbf{L}}$) over all the past domains. 
In the proposed GLRCL framework, $\mathbb{M}_{t}(\cdot)$ learns jointly on the training data from the current domain $D_t$ and latent vectors produced from a pool of generators representing past domains.

\subsubsection{a) Generative latent replay:}
The early layers of deep learning models are responsible for extracting low-level features from input images. After undergoing pre-training on large datasets, these layers' weights become stable and can be effectively reused in various applications, including medical image processing tasks. On the other hand, layers close to the classification head tend to extract discriminative features tailored to specific classes and domains, and fine-tuning them is often essential for maximizing accuracy. We propose to extract rich low-level features from an early layer, coined as generative-replay-layer, and learn generators from these activations. Thus, rather than storing input histopathology samples in an external memory in raw or latent form; we just store the generators and hence avoid possible privacy violation. 
At the arrival of any new domain, separate generators are learned on tumor and healthy images and then the pool containing generators is updated with these newly learned generators. These generators then can be used to produce latent representations mimicking past domains which serve as replay features. 
In each Adam-based mini-batch training, patterns coming from the input layer are concatenated with the replay features at the generative-replay-layer on the mini-batch dimension. 
\[
\arg \min_{\theta_t}   \textit{\textbf{L}} \left(\mathbb{M}_{t (\bar{\theta}, \theta_t)} ((x_t,y_t) \oplus (\mathbb{G}_{t-1} (\mathbf{f}_{t-1}),y_{t-1})\oplus\dots ), \forall t>0\right)
\]
where, $\bar{\theta}$ refers to freezed parameters, $\mathbb{G}_{t-1}(\cdot)$ represents ${(t-1)}^{th}$ domain generator, and $\oplus$ indicates concatenation operation. To uphold the stability of the generated features and the validity of stored generators, we propose freezing all layers beneath the generative-replay-layer and allowing the layers above to learn autonomously, i.e., the forward and backward passes are performed only on layers after generative-replay-layer. Our approach offers applicability to a wide range of encoder models. We use a pre-trained ResNet50~\cite{he2016deep} as our feature extraction model. Then fully connected layers and a classification head are added. 
\subsubsection{b) Domain specific generators:}
For each of the encountered domains having a unique tuple of attributes in set \{stain, organ, center\}, we propose to maintain light-weight generators, separately for each class. The generators are learned from extracted image features and hence cannot be used to produce raw images, ensuring data privacy. Moreover, this opens an avenue for restricted data-sharing in terms of distribution parameter sharing. 
We utilize GMMs as the feature generators. GMMs are well known for their diverse range of data modeling capabilities, also evidenced by their popularity in encoding latent features in deep layers~\cite{bengio2013representation,viroli2019deep}, encoding raw images~\cite{pfulb2021overcoming}, features~\cite{yang2022dynamic,yang2021continual}, etc. Further, they typically require fewer parameters (only mean and variance) compared to any deep neural network-based generators which would otherwise require storing the entire model weights. The cost-effective generators are particularly advantageous, especially in CL scenarios involving multiple distinct domains that require a large number of generators.
Specifically, during training, we extract features from the last batch normalization layers of the ResNet50 model. Then, at the end of each training session, the accumulated features $\mathbf{f}_t \leftarrow BN(x_t)$ are used to learn domain-specific GMMs, for each class. For $t^{th}$ domain and $i^{th}$ class, the generator $\mathbb{G}^i_t(\cdot)$ posses $K$ multivariate Gaussians $\{g_1, g_2 .... g_K\}$ in the mixture with probability density function (PDF) as:
\[
p(\textbf{f}_t) = \sum_{k=1}^{K} w_k \cdot \mathcal{N}(\textbf{f}_t\mid\mu_k, \Sigma_k); \, \,
\mu_k= \frac{1}{N_t} \sum_{n=1}^{N_t} \mathbf{f}_t^n; \, \,
\Sigma_k = \frac{1}{N_t} \sum_{n=1}^{N_t} (\mathbf{f}_t^n - \mu_k)(\mathbf{f}_t^n - \mu_k)^T
\]
where, $N_t$ denotes the number of samples in $t^{th}$ domain for a particular class, $\mathcal{N}(\textbf{f}\mid\mu_k, \Sigma_k)$ represents PDF of each $k^{th}$ Gaussian component $g_k$ with mean $\mu_k$ and covariance matrix $\Sigma_k$, and $w_i$ is a positive weight associated with $k^{th}$ Gaussian such that $\sum_{k=1}^{K} w_k=1$.
For a given value of $K$, the parameters ${\mu_k, \Sigma_k}$ for $k\in \{1,2...K\}$ of mixture distribution are estimated via expectation maximization~\cite{dempster1977maximum}. The fitted mixture model $\mathbb{G}^i_t$ allows to generate random samples reflecting the $i^{th}$ class and $t^{th}$ domain.
Here, the optimal value of $K$ is learned using the Bayesian Information Criterion (BIC)~\cite{fraley1998many}. 
The candidate values for $K$ lie in range 1 and a given upper limit $K_{max}$ , and the one that minimizes BIC is selected.
Finally, the learnt generators from $t^{th}$ domain are then added to GMM-pool containing all past generators $\mathbb{G}^{i}_{t}$ for $i \in\{1,\dots,C\}$ and $t \in \{1,\dots,t-1\}$.

\begin{table}[!b]
\centering
\caption{Details of shift types and available datasets used in experiments}
\label{tab:datasets_samples}
\resizebox{\textwidth}{!}{%
\begin{tabular}{|c|c|c|c|}
\hline
{\bf Details of shift} &
  {\bf Datasets} &
  {\bf \#Train patches} &
  \begin{tabular}[c]{@{}c@{}}{\bf \#Test patches}\end{tabular} \\ \hline

  \begin{tabular}[c]{@{}c@{}}SS: Staining varies, \\organ (breast), center (C1)  \end{tabular} &
  \begin{tabular}[c]{@{}c@{}} HE, CD8, TRI, CK5/14, PAS \end{tabular}&
  \begin{tabular}[c]{@{}c@{}} 1510, 3370,4706, 4704, 3372 \end{tabular}&
  1000 each \\ \hline

\begin{tabular}[c]{@{}c@{}}OS: staining (H), \\organ varies, center (C2),  \end{tabular} &
  \begin{tabular}[c]{@{}c@{}}Breast, Colon, Liver, Lung, Ovary,\\  Pancreas, Prostate, Stomach, Uterus\end{tabular} &
  \begin{tabular}[c]{@{}c@{}}1704, 1982, 1694, 1472, 1338,\\ 1308, 1916, 1836, 1900\end{tabular} &
  600 each \\ \hline

   \begin{tabular}[c]{@{}c@{}}HS: Heterogeneous \end{tabular} &
    \begin{tabular}[c]{@{}c@{}}
  CD8/breast/C1, HE/breast/C3 (BRACS), 
  \\ H/colon/C2,  HE/colon/C4 (CRC)
  \end{tabular}
  &\begin{tabular}[c]{@{}c@{}}3370, 6096, 
  \\1982, 6600 \end{tabular} &
  600-1000  \\ \hline

\end{tabular}
}
\end{table}
\begin{table*}[!htbp]
\centering
\caption{Best performance result in buffer-based / buffer-based with low buffer / buffer-free categories indicated in \textcolor{red}{red} / \textcolor{blue}{blue} / \textcolor{mydarkgreen}{green}, respectively. {\bf Bold:} Upper bound.}
 \label{tab:comp_cl_all}

\begin{tabular}{|c|c|ccc|ccc|ccc|}
\hline
& \begin{tabular}[c]{@{}c@{}}Exp. $\rightarrow$  \end{tabular}
& \multicolumn{3}{c|}{SS}
 & \multicolumn{3}{c|}{OS}
 & \multicolumn{3}{c|}{HS}\\ \hline
 
&Approach 
& \multicolumn{1}{c}{BWT $\uparrow$ } & 
\multicolumn{1}{c}{\begin{tabular}[c]{@{}c@{}} Acc.$\uparrow$\end{tabular}}  &ILM$\uparrow$
& \multicolumn{1}{c}{BWT$\uparrow$} & 
\multicolumn{1}{c}{\begin{tabular}[c]{@{}c@{}} Acc.$\uparrow$\end{tabular}}  &ILM$\uparrow$ 
&\multicolumn{1}{c}{BWT$\uparrow$} & 
\multicolumn{1}{c}{\begin{tabular}[c]{@{}c@{}} Acc.$\uparrow$\end{tabular}}  &ILM$\uparrow$
\\ \hline

\multirow{3}{*}{\rotatebox{90}{Non-CL}} &\begin{tabular}[c]{@{}c@{}}Naive\end{tabular}
&\multicolumn{1}{c}{-31.09}& \multicolumn{1}{c}{71.72}&77.54
&\multicolumn{1}{c}{-7.03} &  \multicolumn{1}{c}{79.41} &82.72 
&  \multicolumn{1}{c}{-38.80}& \multicolumn{1}{c}{66.92}&73.88\\

& Joint 
& \multicolumn{1}{c}{--}& \multicolumn{1}{c}{98.04}&--
&   \multicolumn{1}{c}{} &  \multicolumn{1}{c}{95.46} & --
&  \multicolumn{1}{c}{--}& \multicolumn{1}{c}{86.73}&--\\ 
 	
& Cumulative 
&  \multicolumn{1}{c}{\textbf{-0.30}}& \multicolumn{1}{c}{\textbf{98.00}}&\textbf{98.18}
&\multicolumn{1}{c}{\textbf{8.12}} &  \multicolumn{1}{c}{\textbf{95.80}} &\textbf{94.99 }
&  \multicolumn{1}{c}{\textbf{-0.07}}& \multicolumn{1}{c}{\textbf{88.67}}&\textbf{93.77}\\ \hline

\multirow{6}{*}{\rotatebox{90}{Buffer-based}}&\begin{tabular}[c]{@{}c@{}}GEM \end{tabular} 
&  \multicolumn{1}{c}{-0.83}& \multicolumn{1}{c}{86.64}&83.87
 &\multicolumn{1}{c}{-1.77} &  \multicolumn{1}{c}{88.70} & 89.02
&  \multicolumn{1}{c}{-19.63}& \multicolumn{1}{c}{69.11}&80.93\\

&\begin{tabular}[c]{@{}c@{}}AGEM\end{tabular}
&  \multicolumn{1}{c}{\textcolor{red}{5.63}}& \multicolumn{1}{c}{91.64}&89.83
 &  \multicolumn{1}{c}{-2.18} &  \multicolumn{1}{c}{87.39} &89.38  
&  \multicolumn{1}{c}{-20.07}& \multicolumn{1}{c}{69.31}&80.53\\

&\begin{tabular}[c]{@{}c@{}}ER \end{tabular} 
 &  \multicolumn{1}{c}{-1.62}& \multicolumn{1}{c}{\textcolor{red}{96.80}}&97.10
& \multicolumn{1}{c}{\textcolor{red}{7.83}} &  \multicolumn{1}{c}{\textcolor{red}{\textbf{96.22}}} &\textcolor{red}{ 94.32}&  \multicolumn{1}{c}{-2.60}& \multicolumn{1}{c}{\textcolor{red}{\textbf{90.01}}}&\textcolor{red}{93.53}\\

&\begin{tabular}[c]{@{}c@{}}LR  \end{tabular} 
 &  \multicolumn{1}{c}{-1.06}& \multicolumn{1}{c}{96.72}&\textcolor{red}{97.20}
&  \multicolumn{1}{c}{5.99} &  \multicolumn{1}{c}{94.61} & 93.71
&  \multicolumn{1}{c}{\textcolor{red}{-2.17}}& \multicolumn{1}{c}{87.54}&92.89\\ 
\cline{2-11}

&\begin{tabular}[c]{@{}c@{}}ER*\end{tabular} 
&  \multicolumn{1}{c}{-3.42}& \multicolumn{1}{c}{94.12}&95.87
&  \multicolumn{1}{c}{2.65} &  \multicolumn{1}{c}{\textcolor{blue}{95.50}} & \textcolor{blue}{93.64 }&  \multicolumn{1}{c}{-8.08}& \multicolumn{1}{c}{\textcolor{blue}{89.07}}&\textcolor{blue}{91.94}\\ 

&\begin{tabular}[c]{@{}c@{}}LR*\end{tabular}
&  \multicolumn{1}{c}{\textcolor{blue}{-2.41}}& \multicolumn{1}{c}{\textcolor{blue}{95.28}}&\textcolor{blue}{95.97}
 &  \multicolumn{1}{c}{\textcolor{blue}{5.97}} &  \multicolumn{1}{c}{93.76} & 92.87 
 &  \multicolumn{1}{c}{\textcolor{blue}{-5.02}}& \multicolumn{1}{c}{85.97}&91.54\\ \hline

\multirow{4}{*}{\rotatebox{90}{Buffer-free}} & \begin{tabular}[c]{@{}c@{}}SI \end{tabular}
&  \multicolumn{1}{c}{-26.86}& \multicolumn{1}{c}{78.56}&79.83
 &  \multicolumn{1}{c}{-6.77} &  \multicolumn{1}{c}{82.31} &84.51 
 &  \multicolumn{1}{c}{-35.92}& \multicolumn{1}{c}{70.71}&75.45\\

&\begin{tabular}[c]{@{}c@{}}LwF \end{tabular} 
&  \multicolumn{1}{c}{-28.99}& \multicolumn{1}{c}{70.00}&78.28
 &  \multicolumn{1}{c}{-4.15} &  \multicolumn{1}{c}{83.44} &84.45 
 &  \multicolumn{1}{c}{-30.32}& \multicolumn{1}{c}{74.72}&77.50\\

&\begin{tabular}[c]{@{}c@{}}EWC \end{tabular}
&  \multicolumn{1}{c}{-27.16}& \multicolumn{1}{c}{77.56}&80.15
 &   \multicolumn{1}{c}{-9.12} &  \multicolumn{1}{c}{81.00} &82.69  
 &  \multicolumn{1}{c}{-39.42}& \multicolumn{1}{c}{66.41}&73.54\\ 

&\begin{tabular}[c]{@{}c@{}}Proposed\end{tabular}
&  \multicolumn{1}{c}{\textcolor{mydarkgreen}{-1.65}}& \multicolumn{1}{c}{\textcolor{mydarkgreen}{96.20}}&\textcolor{mydarkgreen}{96.76}
&  \multicolumn{1}{c}{\textcolor{mydarkgreen}{6.43}} &  \multicolumn{1}{c}{\textcolor{mydarkgreen}{94.48}} &\textcolor{mydarkgreen}{93.23} & 
\multicolumn{1}{c}{\textcolor{mydarkgreen}{-2.58}}&\multicolumn{1}{c}{\textcolor{mydarkgreen}{87.38}} & \textcolor{mydarkgreen}{92.66}\\ 
\hline

\end{tabular}%
\end{table*}

\section{Experimental Setup}
\subsubsection{(1) Datasets:}
Various datasets characterizing shifts in staining, organ, and center are curated. We outline these datasets in Table~\ref{tab:datasets_samples}. Please note that train-test split does not follow slide-level split except for the organ shift datasets.

\textbf{Stain shift (SS):}
The data was acquired at center C1 and includes healthy and tumorous breast tissue. Experienced pathologists carefully annotated breast cancer regions in detail for Hematoxylin and eosin (HE), Cytokeratin (CK5/14), and Cluster of differentiation (CD8) stained WSIs. For preprocessing, we selected foreground tissue regions using the CLAM preprocessing pipeline~\cite{clam_prepro}. As stainings highly vary in colorization, we manually adjusted tissue detection thresholds. Each annotated region was tiled into $512\times512$ px patches at $40\times$ magnification. Since we obtained $5\times$ more patches from CK5/14 and CD8 compared to HE and to include more inter-stain variation, we used half of both datasets and augmented them using Cycle-GAN (cGAN) into artificial trichome (TRI) and periodic acid-Schiff (PAS) stainings, respectively. 

\textbf{Organ shift (OS):}
This dataset from center C2 comprises 20 tissue microarray (TMA) cores, occupied with tissue of 9 organs in hematoxylin (H) staining. TMA cores were systematically sampled from a diverse cohort of patients and contained cancerous and healthy tissue. The same preprocessing strategy as in stain-variation is applied here. We split datasets spot-wise to prevent data leakage at both, patient and tissue level. 

\textbf{Heterogeneous shift (HS):} In addition to homogeneous kinds of shift (i.e., either SS or OS), we also consider heterogeneous shifts involving a mix of different stains, organs, and centers, simultaneously. Specifically, we consider the "CD8" dataset from the SS sequence (stain:CD8, organ:breast, center:C1), the "Colon" dataset from the OS sequence (stain:H, organ:colon, center:C2), a subset from the BRACS dataset~\cite{brancati2022bracs} (stain:HE, organ:breast, center:C3), and a subset from the CRC dataset~\cite{kather2019predicting} (stain:HE, organ:colon, center:C4). All these datasets are curated such that they comprise only normal and tumor classes.

\subsubsection{(2) Comparable Methods:}
Both non-CL and CL approaches are considered for comparison. The non-CL approaches include (a) lower-bound performance obtained by a naive method (i.e. training dataset of the current task is used only) and (b) upper-bound performance by a joint method (i.e. training datasets from all tasks are used) and a cumulative method (training datasets are accumulated from seen tasks). For CL, we consider frequently used approaches in the two categories as (a) buffer-based (i.e. a few past samples are stored in raw or latent form) and (b) buffer-free (no storing of past samples). In the buffer-based category, we consider GEM~\cite{lopez2017gradient}, A-GEM~\cite{chaudhry2018efficient}, ER~\cite{rolnick2019experience}, and LR~\cite{pellegrini2020latent}. In the buffer-free category, EWC~\cite{kirkpatrick2017overcoming}, SI~\cite{zenke2017continual}, and LwF~\cite{radio3} are considered.

\subsubsection{(3) Implementation details:} In the following, we describe the training strategies of our models and the included evaluation metrics.

\textbf{cGAN pre-training:} For cGAN pre-training we extracted patches in the scale of $512\times512$ px from TRI and PAS stained WSIs as provided by the KPMP database~\cite{kpmp}. Regarding hyperparameter settings, we followed the literature~\cite{nassim} and performed a grid search around an order of magnitude around the base parameters. We selected the best-performing model by visually inspecting the derived stain augmentations, resulting in the number of iterations of $30\in \left[20, 30, 40\right]$ epochs and a learning rate of $1.5e-4 \in \left[1.0e-3, 1.5e-3, \dots, 1e-5\right]$. To prevent any data leakage, we used unlabeled excess tiles from CK5/14 and CD8 from our stain-variation WSIs to train our cGAN using target domains from the KPMP dataset. A total of 5000 images were used for training. The training and inference time was approximately 8 hours per stain.

\textbf{CL training:} We selected the hyperparameters according to common CL evaluation schemes. For all comparable methods, we used the CL benchmark library Avalanche\footnote{https://avalanche.continualai.org/}. We used AdamW as an optimizer and used learning of $1e-03$. Images were resized to $256\times256$ px and batched with a size of 64. All methods used a pre-trained ResNet50 model initialized with ImageNet weights. After the last batch normalization layer in ResNet50, 5 fully connected layers \{512, 256, 128, 64, 32\} are added. For all the experiments, we keep $K_{max}=10$. All experiments were conducted using an Nvidia A100 GPU, with training sessions typically completed within a maximum duration of 4 hours across all approaches.

\textbf{Evaluation metrics:} The performance is assessed using the classification accuracy metric. For any experiment with $T$ tasks in sequence, we get a train-test matrix $A$ of size $T\times T$ where the cell value $A_{ij}$ represents accuracy on $j^{th}$ task after training up to $i^{th}$ task. This matrix can be used to compute various metrics to compare approaches. We consider popularly used metrics in CL literature, specifically Backward Transfer (BWT) for measuring forgetting (Eq. 3 in~\cite{diaz2018don}), Incremental Learning Metric (ILM) to measure incremental learning capability (Eq. 2 in~\cite{diaz2018don}) and average accuracy across all the tasks after learning the last task (Eq. 2 in \cite{lopez2017gradient}). For all these metrics a higher positive value indicates superiority.

\begin{figure}[!ht]
    \centering
     \includegraphics[scale=0.38]{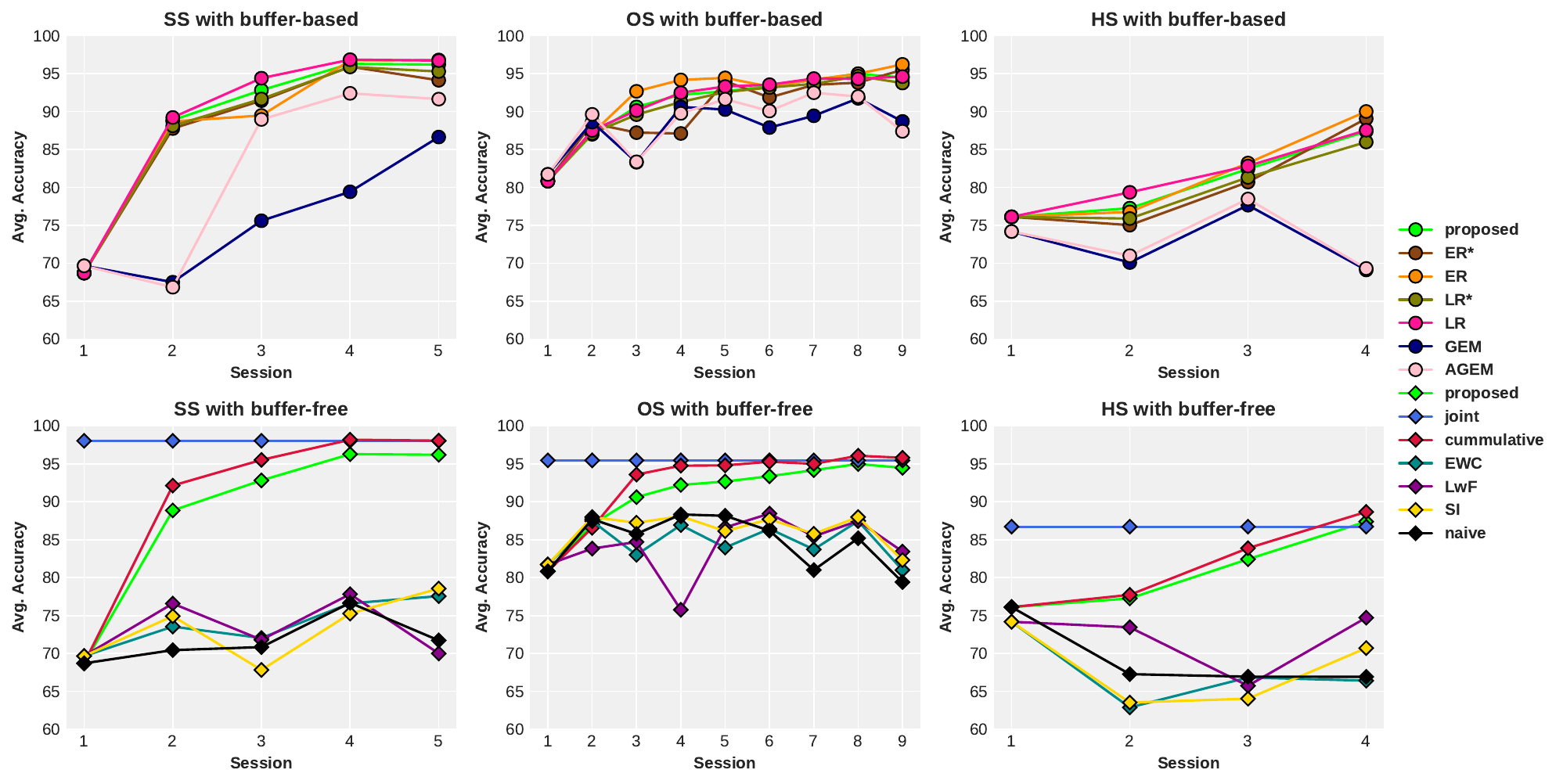}
\caption{Performance over time for different shifts (SS, OS, HS) and for buffer-based {\it (upper row)} versus buffer-free CL approaches + non-CL {\it (lower row)}.}
\label{fig:figSeq1AllShift}
\end{figure}
\section{Results and Discussion}

Within each shift scenario (SS, OS, HS), a random ordering of the available data is created and utilized for evaluating different approaches (see Table~\ref{tab:comp_cl_all}). Our supplementary file provides results on further random orderings (showing same trends) and further discussion on individual evaluation metrics. As expected, "cumulative" and "joint" represent the upper, and "naive" indicates the lower bound. For all three evaluation metrics, there is a large performance gap between buffer-based approaches and buffer-free approaches across all experiments (SS, OS, and HS). This is in line with previous findings that rehearsal-based approaches handle forgetting in medical data better~\cite{kumari2023continual,derakhshani2022lifelonger}. 
In the buffer-based category, ER and LR approaches largely offer the best results. Our buffer-free GLRCL approach (green) achieves performances slightly worse than the best-performing buffer-based approaches with a large buffer (red).
However, when reducing the buffer size by 1/4 (see ER* and LR*), a significant performance drop is observed (best results in blue), resulting in performance values mostly below our GLRCL approach in all experiments. This shows that the performance of buffer-based approaches gets compromised when reducing the buffer size. When comparing with existing buffer-free CL approaches, we observe that our novel GLRCL method clearly outperforms these methods by a large margin across all three domain shift conditions.
Further, in terms of accuracy (Acc. and ILM), HS performs generally worse than SS and OS, due to the multitude of variations (stain, organ, center shifts) encountered in HS.

To analyze performance over time, we report the average accuracy of the cancer classification tasks after each training session for each of our three domain shift experiments (see Fig~\ref{fig:figSeq1AllShift}). The first row shows a comparison of our GLRCL approach (green, "proposed") with buffer-based, and the second row with buffer-free and non-CL approaches, respectively. These graphs clearly show that GLRCL significantly outperforms the buffer-free approaches and performs similarly to the buffer-based approaches, creating a new CL benchmark and an alternative to buffer-based CL approaches for digital pathology.

\section{Conclusion}\label{sec:conc}
Our novel privacy-aware CL approach for histopathology tumor detection in domain incremental scenarios outperforms existing buffer-free CL approaches by a large margin, and mitigates data privacy concerns compared to image-storing buffer-based CL methods with comparable performance. Our work significantly advances privacy-aware tumor detection from histology data, pushing the boundaries of current CL strategies under privacy preservation. For future work, we investigate for a better choice of latent representations so that the learned GMM better captures the domain and class differences. Additionally, we aim for a single generator with a dynamic update mechanism triggered with drift detection module which eventually deals with reoccurring and overlapping domains.

\subsubsection{Acknowledgements}
This work was supported by the German Research Foundation (Deutsche Forschungsgemeinschaf, DFG) under project number 445703531 and the German Federal Ministry of Education and Research (Bundesministerium für Bildung und Forschung, BMBF) under project numbers 01IS21067A and 01IS21067B. The authors gratefully acknowledge the computational and data resources provided by the Leibniz Supercomputing Centre (www.lrz.de).

\subsubsection{Disclosure of Interests}
The authors declare that they have no conflicts of interest related to this work.

\clearpage

\bibliographystyle{splncs04}
\bibliography{main}

\end{document}